\documentstyle[epsbox,twocolumn]{article}
\title{Self-Similar Traffic Originating in the Transport Layer}
\author{Kensuke Fukuda\\[0.2cm]
Network Innovation Laboratories,\\ 
Nippon Telegraph and Telephone Corp.\\
3-9-11 Midori-cho Musashino, \\180-8585, Japan\\
fukuda@t.onlab.ntt.co.jp
\and Misako Takayasu\\[0.2cm]
Faculty of Complex Systems, \\Future University-Hakodate\\
116-2 Kameda-Nakano, Hakodate,\\
041-0803, Japan\\
takayasu@fun.ac.jp
\and Hideki Takayasu\\[0.2cm]
Sony Computer Science Laboratories\\
3-14-13 Higashi-Gotanda, Shinagawa\\
141-0022, Japan\\
takayasu@csl.sony.co.jp}
\date{}
\begin{document}
\thispagestyle{empty}
\maketitle
\thispagestyle{empty}
\begin{abstract}
We performed a network traffic simulation  to clarify the  
mechanism producing self-similar traffic originating in the 
transport layer level.
Self-similar behavior could be observed without assuming a 
long-tailed distribution of the input file size.
By repeating simulations with modified TCP we 
found that the feedback mechanism from the network, such as  
packet transmission driven by acknowledgement packets, 
plays an essential role in explaining the self-similarity 
observed in the actual traffic. 
\end{abstract}
\section{INTRODUCTION}

Internet traffic fluctuation is known to show  
self-similarity or long-range dependency
\cite{Leland94,Csabai94,Paxson95,Crovella97,Willinger97}. 
This self-similarity is the scale invariant property that 
the burst size of the flow density fluctuation 
seems to have the same tendency at various observation time scales. 
Recently, it was pointed out that 
network traffic behavior can be regarded as phase transition 
phenomena in statistical physics\cite{Takayasu96a,Takayasu99a,Fukudaphd}, 
which naturally involves the self-similar model.
The phase transition\cite{Stanley71} is characterized by dynamical 
phase changes between non-congested and congested phases, and 
self-similarity can be observed at the critical point between these two phases. 

Similar to the observations, some studies investigated the mechanism 
of the self-similarity observed in network traffic. 
In the application layer level, 
Crovella et al. explained self-similar traffic from the viewpoint that 
the sizes of files on the web server have a power-law 
distribution\cite{Crovella97}. 
For the datalink layer, Fukuda et al. showed that 
self-similar Ethernet traffic can be reproduced by the 
effects of the contention between the nodes and of  
the exponential backoff mechanism at the packet collisions\cite{Fukuda2000a}. 
Also, 
Park et al.\cite{Park97} and Feldmann et al.\cite{Feldmann99} 
pointed out that the transport layer functionality (especially TCP) strengthens 
the long-range dependency. 
Although they aimed to show the transport layer effect, 
their approaches implicitly assume the long-range dependency in the 
application level such  as the file size distribution of the application. 
Thus, there seems to be little understanding of the physical explanation 
of the transport layer functionality itself from the viewpoint of the 
generation of self-similar traffic. 
 
In this paper, we focus on the transport layer effect 
independent of the application level causality. 
In other words, we investigate essential factors for generating  
self-similar traffic in TCP. 
To clarify them, 
we performed a simple topological simulation using the ns-2 simulator. 
The results show that phase transition phenomena (and 
self-similarity) quite similar to those observed 
in the actual traffic can occur in 
aggregated TCP traffic even when   
input traffic sources have no temporal correlation or  
long-range distribution in  file size. 
We also demonstrate that the feedback mechanism (especially 
acknowledgement packet driven events) plays  
an important role in generating the self-similarity observed in  
actual traffic, though the rate control and retransmission mechanism have less 
impact on it.

\section{SIMULATION SETUP AND ANALYSIS METHOD}
\subsection{Simulation Setup}
In this simulation, we used the VINT network simulator ns-2, and 
added some tcl scripts and C++ code. 
The scenario was file transfer from the sender to the receiver on the 
leaf nodes. 
Figure 1 shows our simple simulation topology consisting of  
three leaf nodes and one router. 
\begin{figure}[htbp]
\begin{center}
\epsfile{file=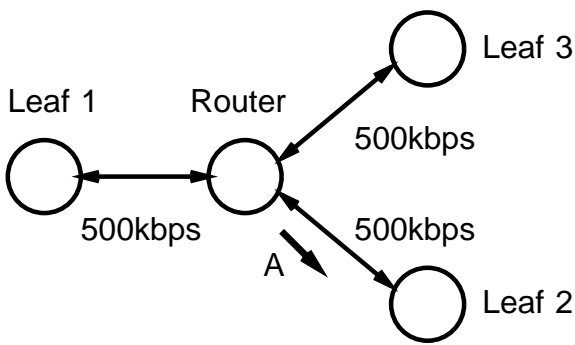,scale=0.9}
\end{center}
\caption{Network topology.}
\end{figure}
A connection employed the TCP Reno as the basic transport protocol, and 
a TCP connection was established between a pair of randomly selected 
leaf nodes. 
The connection interval times of the connection were exponentially distributed. 
It is important to note that the number of packets in a connection  
followed an exponential distribution (mean size = 100 packets) 
throughout this simulation. 
Namely, the distribution of the number of packets 
had no temporal correlation, which means that there was no 
application-level causality. 
This condition is needed to clarify the transport mechanism 
from the viewpoint of the self-similarity. 
The buffer sizes in the leaf nodes and the router were set to 800 packets in 
most simulation, and the packet size was set to 576 bytes. 
Also, the bandwidth and transmission delay of each half duplex link were  
$500$ kbyte/sec and $50$ msec., respectively. 
The large buffer size was chosen for easy analysis of the fluctuation of  
packets in the buffer. 
Our results were obtained from several runs of the simulation, 
each lasting $4800$ seconds. 
We were interested in the statistical behaviors of the aggregated traffic 
streams when the connection arrival rate (to be denoted as ``r'') 
to the leaf nodes varied. 

\subsection{Analysis Method}
In order to examine the self-similar nature of network traffic fluctuation, 
we focused on two empirical distributions, 
namely the congestion duration and the recurrent time of the queue length.

The congestion duration length distribution is a well-known distribution  
for judging the self-similarity of a given time series\cite{Takayasu93b,Willinger97}. 
The congestion state is defined by the condition that 
the output flow density in the observed link is larger than a certain 
threshold flow density. 
Then, the congestion duration length (L) is calculated as  
the sequential number of  congestion states multiplied by 
the bin size.
The cumulative distribution of this duration ($P(>L)$)  
is a power-law distribution with exponent -1.0 
($P(>L) \propto L^{-1.0}$) when the original flow fluctuation is 
self-similar\cite{Takayasu93b}, 
which is characterized by the $1/f$ type power spectrum.
Theoretically, the power-law distribution is observed independent of 
the value of the threshold if the original fluctuation is self-similar.

The recurrent time of the queue length is 
introduced as the duration time until the queue length becomes zero. 
The cumulative distribution of such recurrent time obeys the same 
power-law distribution with exponent $-1.0$ as the congestion duration length.

We checked the congestion duration length of the link from 
the router to leaf node 2 in Figure 1 (denoted by A). 
Also, we observed the queue length at the Router's output queue to Leaf node 3.

\section{TRANSPORT LAYER EFFECT}
\subsection{Real Traffic Flow}
In this subsection, we review the traffic fluctuation 
in an actual network focusing on the self-similarity and the phase 
transition phenomena. 

\begin{figure}[htbp]
\begin{center}
\epsfile{file=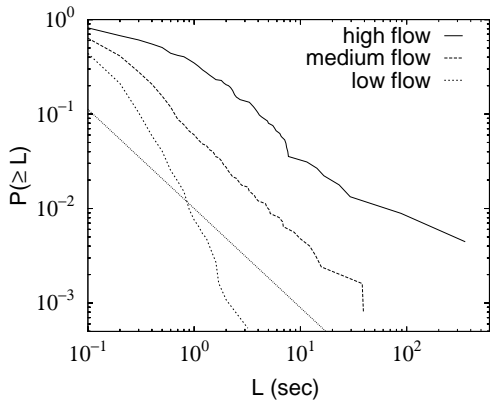,scale=0.6}
\end{center}
\caption{Congestion duration length in actual network traffic. 
The straight line indicates the slope $-1.0$.}
\end{figure}
Figure 2 shows the cumulative distribution of the 
congestion duration length in an actual traffic flow\footnote{
More detailed analysis is shown in  \cite{Fukudaphd}.}. 
This traffic trace was collected in the 
Ethernet link connecting the WIDE backbone in Japan and 
the US west coast for 4 hours; 
80\% of the traffic was due to web applications. 
The three curves in the figure indicate the difference in  mean flow 
density of the traffic flow. 
This figure shows that for medium flow 
the distribution is approximately the power-law distribution with 
exponent $-1.0$ representing  self-similarity.
However, away from the critical point the distributions deviate from the 
power-law distribution. 
When the total amount of traffic is small, 
the congestion duration length obeys an exponential distribution 
characterized by the short-time dependency.  
On the other hand, 
the larger traffic density case denoted by the high flow in this figure 
demonstrates the existence of the large-cluster congestion. 
Consequently, this result clearly shows that self-similarity occurs in  
a special case in the actual network traffic, and the phase transition view 
can capture all the properties\cite{Fukudaphd,Takayasu99a}.

\subsection{TCP Traffic Behavior}
We are interested in the mechanism of generating the self-similar traffic 
observed in the previous subsection from the standpoint of the transport 
layer functionality.
This subsection explains a numerical simulation with an ordiary TCP Reno 
algorithm. 

\begin{figure}[htbp]
\begin{center}
\epsfile{file=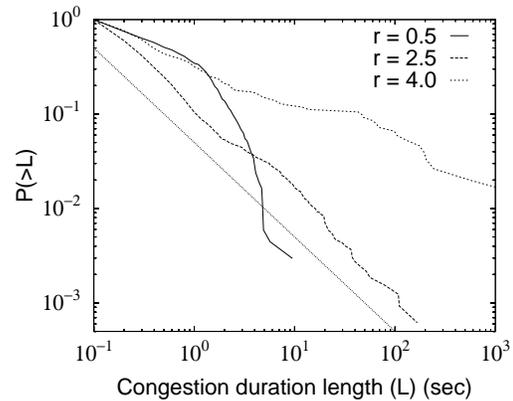,scale=0.6}
\end{center}
\caption{Congestion duration length of TCP Reno. 
The straight line indicates the slope $-1.0$.}
\end{figure}
Figure 3  shows the cumulative distribution of the 
congestion duration length of the aggregated TCP traffic flows at  link A 
in Figure 1. 
The threshold value of congestion level was empirically set to 
5000 bytes throughout this simulation. 
The three lines correspond to three different connection interval rates
(r = 0.5, 2.5, 4.0 connections/sec). 
The mean connection duration times were 1.53, 13.4, and 412.64 sec., 
respectively. 
We found that the slope of the distribution at r = 2.5 was approximately 
$-1.0$, which is recognized as the critical point behavior (
the slope of the straight line is $-1.0$ in the figure).
For medium  connection arrival rate 
the traffic flow had self-similarity like  the actual traffic. 
Also, the plot decays exponentially below the critical point (r = 0.5), 
and it was characterized by a stretched curve above the critical point
(r = 4.0) representing the existence of coarse-grained congestion. 
These curves are completely consistent with the actual traffic 
behavior. 
The most significant point in this simulation is  that 
this power-law distribution with exponent $-1.0$ could be observed 
even when the input traffic 
followed an exponential distribution, not assuming a fat-tail distribution. 
Namely, the self-similarity appeared in the traffic fluctuation  
independent of the input file size distribution at the critical point.

\begin{figure}[htbp]
\begin{center}
\epsfile{file=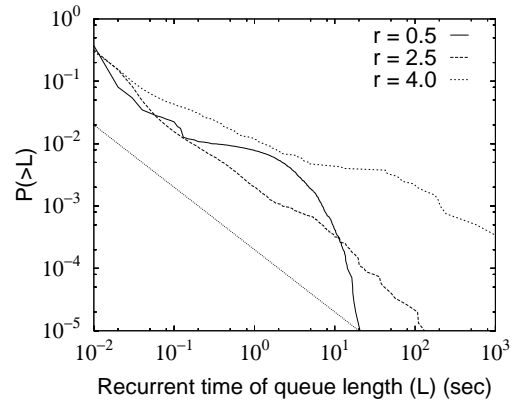,scale=0.6}
\end{center}
\caption{Recurrent time of queue length. 
The straight line indicates the slope $-1.0$.}
\end{figure}
Next, we show the distribution of the recurrent time of the queue length 
at the router's queue (Figure 4). 
Again, the plotted curve  followed the 
power-law distribution with exponent -1.0 at the critical point (r = 2.5) 
which is the same connection arrival rate as in the congestion duration 
length analysis.
Also, when the connection arrival rate was smaller, the 
plotted curve exhibited a quick decay, and 
a larger rate led to a more stretched distribution due to 
the large size congestion.  

Additionally, we confirmed that the congestion duration length distribution 
and the recurrent time distribution had the same 
phase transition behavior in all  leaf nodes (links).

\subsection{Effect of TCP Component}
Section 3.2 clarified that 
transport functionality (TCP) itself plays a role in producing 
self-similar traffic. Now we 
focus on the generation mechanism of the self-similarity in 
the aggregated traffic fluctuation.
This subsection explains 
the results of additional simulations based on modified TCP 
in order to clarify the mechanism of the 
power-law distribution with exponent $-1.0$ observed in the previous subsection.

The first modification is not to use the slow start algorithm which increases 
the transmission rate exponentially. 
The modified algorithm employs a linear rate increment even in 
the connection starting time, instead of the original exponential rate increment.

Figure 6  shows the congestion duration length distribution of 
the linear rate increment case with feed back control. 
The aggregated traffic also exhibited phase transition similar to the normal 
TCP simulation. 
Thus, the phase transition and corresponding self-similarity are 
known to be independent of the details of the increment algorithm of 
the transmission rate.
\begin{figure}[htbp]
\begin{center}
\epsfile{file=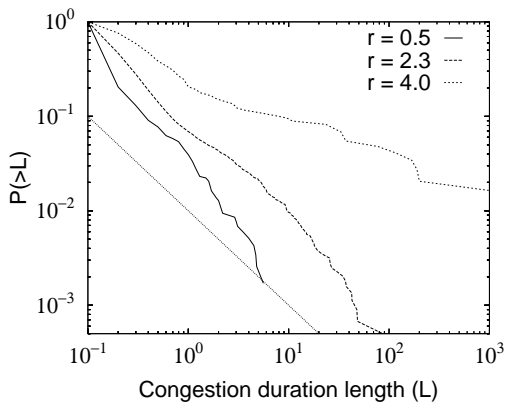,scale=0.6}
\end{center}
\caption{Congestion duration length of linear start TCP. 
The straight line indicates the slope $-1.0$.}
\end{figure}

Next, we modified the feedback control algorithm. 
We checked two non-feedback control schemes, CBR over UDP, and 
a linear rate increment algorithm over UDP. 
In CBR-over-UDP simulation, the packet inter-arrival duration time  
was set to 20 msec. 
Also, the linear rate increment algorithm includes a  method  
in which the transmission rate from the sender increases by 1 
for the fixed interval (150 msec).
The latter method is similar to the previous linear increment algorithm 
modified from  the original TCP. 
The difference is in the trigger of the packet transmission event; 
namely, the packet transmission event is based on the fixed-time interval 
event independent of the reception of the acknowledgement packets.
The distributions of the number of 
packets to be sent and connection arrival duration are exponential 
as in the previous simulations.

The distribution of congestion duration length for non-feedback 
control algorithm is shown in Figure 6. 
The two non-feedback algorithms reproduce similar statistical tendencies of 
the congestion duration length, therefore, we only show the result of the 
linear rate increment algorithm over UDP. 
We found that the exponent of the power-law distribution had a value, $-0.5$, 
clearly different from the exponent, $-1.0$, obtained in the previous 
subsection, 
although phase transition behavior is observed quantitatively like the 
previous simulations (the straight line in the figure indicates slope $-0.5$).
This type of exponent is known for the single buffer system with 
Poisson input\cite{Takayasu96a}. 
Thus, from this simulation, we can conclude that the feedback mechanism is 
important in generating a self-similar fluctuation with  exponent $-1.0$,  
which is observed in  actual Internet traffic.
\begin{figure}[htbp]
\begin{center}
\epsfile{file=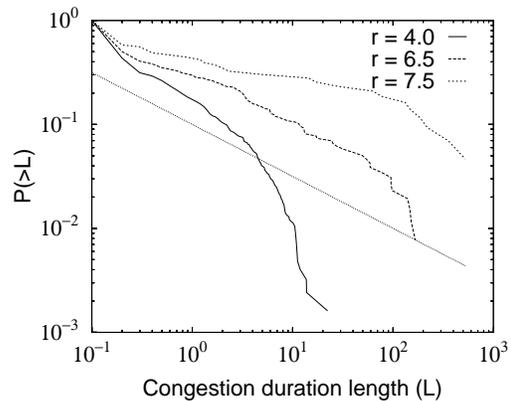,scale=0.6}
\end{center}
\caption{Congestion duration length of non-feedback algorithm. 
The straight line indicates the slope $-0.5$.}
\end{figure}

\subsection{Effect of Buffer Capacity}
\begin{figure}[htbp]
\begin{center}
\epsfile{file=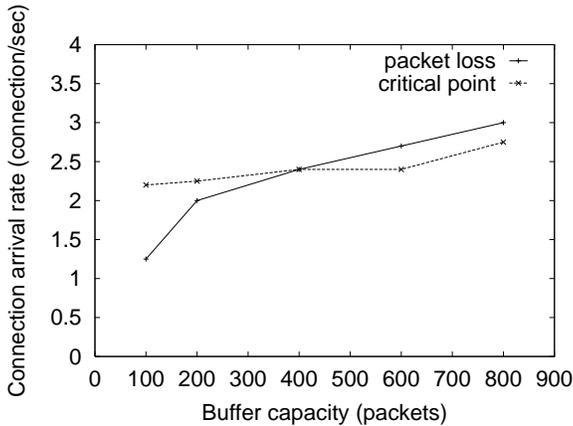,scale=0.6}
\end{center}
\caption{Packet loss and critical point vs. buffer capacity.}
\end{figure}
Figure 7 plots the connection arrival rate at which  
packet loss is first observed in the system 
as a function of the buffer capacity in the nodes 
together with the critical connection arrival rate showing the power-law 
distribution.

This figure shows that both the critical point and the packet 
loss point become larger as  
the buffer capacity increases.  
However, it should be emphasized that self-similarity was observed without 
packet loss event for buffer capacity larger than 400, 
namely, the retransmission event is not directly 
concerned with the generation of the phase transition. 
Moreover, we confirmed that there was no timeout event of the retransmission 
timer in above range. 
It is an evidence that the exponent of the power-law distribution 
is independent of the method of the retransmission.
Also, the larger buffer capacity leads to a larger critical point value,  
however, the buffer capacity itself does not 
affect the generation of the phase transition phenomena. 

\section{CONCLUDING REMARKS}
In this paper we  focused on a simple mechanism for generating  
the self-similarity observed in actual network traffic at 
the transport layer. 
In order to clarify the essence of the mechanism, 
we performed simulations with simple settings such as 
exponential file size, fixed-size packets, and the simple topology.

We showed that the self-similar traffic is observable in these simple 
settings when the traffic sources send packets using the normal TCP algorithm.
In addition, the reproduced traffic behavior is consistent with the 
phase transition model. 
The most significant result is that the self-similarity appears even with the 
exponential input file size. 
This indicates that the transport protocol itself includes the 
mechanism generating self-similar traffic. 

Moreover, we  clarified that the 
feedback mechanism, especially the packet transmission triggered by 
the acknowledgement packet, in TCP 
is an essential factor in generating the 
self-similarity from the results for the modified algorithm.
Traffic employing non-feedback effect with linear rate incremental algorithm 
 exhibits similar phenomena, 
however, the exponent of the power-law distribution, $-0.5$, is 
inconsistent with that of the TCP with linear rate increment algorithm, $-1.0$. 
This indicates that an essential factor is the acknowledgement-driven packet 
transmission rather than the timer-driven one.
We also confirmed in both the normal and modified TCP simulation that 
the distribution of the inter-packet arrival of the acknowledgement packet in a TCP 
connection has a power-law distribution at the critical point
(between 0.01 -- 1.0 seconds).
Namely, the origin of the self-similar traffic is likely due to 
the feedback mechanism from the network such as the delay of  
acknowledgement packets.

Finally, we showed that the retransmission mechanism and 
buffer capacity have less impact on the generation of self-similarity 
(1/f type fluctuation).
We concluded that these effects only work to stretch the time spent staying 
in the self-similar state. 
The network state seems to have been heavily congested at the 
critical connection arrival rate in our simulation. However, 
there are congested routers in actual wide area networks, 
and our results indicate that if a flow passes through the router in the 
critical state once, the flow can be self-similar. 

Our simulation results have extracted the essence of the origin of  
self-similar traffic from an actual complex network system 
(both topologically and algorithmically). 
The future direction of this research will be to support the development of  a 
more effective congestion control algorithm based on the knowledge 
obtained from these simulations. 

\section*{ACKNOWLEDGEMENTS}
We wish to thank Kenjiro Cho and Toshio Hirotsu for helpful discussion. 

\bibliographystyle{plain}

\section*{AUTHOR BIOGRAPHIES}
\noindent{\bf KENSUKE FUKUDA} is a research scientist in Network Innovation 
Laboratories at Nippon Telegraph and Telephone Corpration (NTT). 
He received a B.S. in Electrical Engineering, M.S.  
and Ph.D. in Computer Science in Keio University, Japan. 
His research interests include network traffic analysis and modeling, 
traffic control, and network protocol design. 

\noindent{\bf MISAKO TAKAYASU} is an associate professor in 
department of complex systems, Future University - Hakodate, Japan.  
She received a B.S. in Physics at Nagoya University,
and Ph.D. in Science at Kobe University, Japan.
Her research interests include non-equilibrium physics, especially in
dynamic phase transition, and information physics.

\noindent{\bf HIDEKI TAKAYASU} is a senior researcher 
at Sony Computer Science Laboratories Inc.
He received a B.S. and Ph.D. in Physics at Nagoya University. He was a
professor in Graduate School of Information Sciences at 
Tohoku University, Japan. His research covers wide area of
topics relating to fractals, such as stable distributions, earthquakes and
econophysics.

\end{document}